\documentclass[onecolumn]{aa} 

\usepackage{graphicx}
\usepackage{threeparttable}
\usepackage{txfonts}
\usepackage[allcolors=blue]{hyperref}
\begin{document} 

   \title{A long-lived compact jet in the black hole X-ray binary candidate AT2019wey}


   \author{Hong-Min Cao\inst{1}
          \and
          Giulia Migliori\inst{2}
          \and
          Marcello Giroletti\inst{2}
          \and
          S\'{a}ndor Frey\inst{3, 4}
          \and
          Jun Yang\inst{5}
          \and
          Krisztina \'E. Gab\'anyi\inst{6, 7, 3}
          \and
          Lang Cui\inst{8}
          \and
          Tao An\inst{9}
          \and
          Xiao-Yu Hong\inst{9}
          \and
          Wen-Da Zhang\inst{10}
          }
   
   \institute{School of Electronic and Electrical Engineering, Shangqiu Normal University, 298 Wenhua Road, Shangqiu, Henan 476000, China\\
            \email{hongmin.cao@foxmail.com}
         \and
             INAF - Istituto di Radioastronomia, Via Gobetti 101, I-40129, Bologna, Italy
         \and
             Konkoly Observatory, ELKH Research Centre for Astronomy and Earth Sciences, Konkoly Thege Mikl\'os \'ut 15-17, H-1121 Budapest,
             Hungary
         \and
            Institute of Physics, ELTE E\"otv\"os Lor\'and University, P\'azm\'any P\'eter s\'et\'any 1/A, H-1117 Budapest, Hungary
         \and
            Department of Space, Earth and Environment, Chalmers University of Technology, Onsala Space Observatory, SE-43992 Onsala, Sweden
         \and
            ELTE E\"otv\"os Lor\'and University, Institute of Geography and Earth Sciences, Department of Astronomy, P\'azm\'any P\'eter s\'et\'any 1/A, H-1117 Budapest, Hungary
         \and
            ELKH-ELTE Extragalactic Astrophysics Research Group, E\"otv\"os Lor\'and University, P\'azm\'any P\'eter s\'et\'any 1/A, H-1117 Budapest, Hungary
         \and
            Xinjiang Astronomical Observatory, Chinese Academy of Sciences, 150 Science 1-Street, Urumqi, Xinjiang 830011, China
         \and
            Shanghai Astronomical Observatory, Key Laboratory of Radio Astronomy, Chinese Academy of Sciences, 80 Nandan Road, Shanghai 200030, China
         \and
            National Astronomical Observatories, Chinese Academy of Sciences, 20A Datun Road, Chaoyang District, Beijing 100012, China
           }

   \date{Received Xxxx YY, 2021; accepted Xxxx YY, 2021}
 
  \abstract
   {AT2019wey is a transient discovered by the Asteroid Terrestrial-impact Last Alert System (ATLAS) survey in December of 2019. Follow-up optical, radio, and X-ray observations led to classification of this source as a Galactic black hole X-ray binary candidate. We carried out one-epoch 6.7 GHz European VLBI (Very Long Baseline Interferometry) Network (EVN) and two-epoch multi-frequency (1.6, 4.5, 6.7~GHz) Very Long Baseline Array (VLBA) observations within a year after its discovery. These observations reveal a fading and flat-spectrum radio source with no discernible motion. These features suggest the detection of a compact jet. The source appears resolved at milliarcsecond scales, and the source angular size versus frequency trend is consistent with scatter broadening. This allows us to constrain the lower limit of the source distance to 6~kpc if the scattering medium is in a Galactic spiral arm. 
   For a source location at greater than 3 kpc, the estimated upper limit of the peculiar velocity suggests the asymmetric natal kick may have occurred during the black hole formation stage.}


   \maketitle
%

\section{Introduction}

\label{sec-1}
   
   Galactic black hole X-ray binaries (BHXBs) are stellar-mass black holes fed by gaseous material from their companion stars via Roche-lobe outflow for the low-mass BHXBs or strong stellar wind for the high-mass ones \citep[e.g.][]{corral16}. 
   BHXBs spend most of their time in the quiescent states, and manifest themselves as transient sources when they enter a state of outburst lasting days to years. In an outburst state, based on the X-ray spectral and timing properties, BHXBs usually evolve from the low/hard state (LHS) to high/soft state (HSS), and then return to LHS. A compact jet is observed in the LHS, while the radio core activity is quenched in the HSS. During the transition from LHS to HSS, a transient jet is often observed \citep[see][for a reference]{fender04,fender09}. The short timescales of the outbursts facilitate multi-wavelength photometric, spectral, and imaging monitoring, allowing us to quasi-simultaneously track changes of the disk--corona structure and the evolution of the radio jet activity. The latter can be directly resolved by the technique of very long baseline interferometry (VLBI) and/or connected-element interferometers \citep[e.g.][]{hjellming95,tingay95,hannikainen01,miller-Jones12,paragi13,bright20}. Galactic BHXBs are miniature versions of extragalactic active galactic nuclei powered by accretion onto supermassive ($\sim 10^{6}-10^{10}\, \mathrm{M}_{\odot}$) black holes. The multi-band observing efforts could provide critical insights for unveiling the jet formation mechanism across all mass scales of the accreting black hole systems, which is still not fully understood.
   
   Multi-epoch VLBI observations could be used to accurately pinpoint the radio core of the compact jet on the sky via relative astrometry, and thus measure its parallax and proper motion over longer timescales. The distance can then be inferred or constrained. If a radial velocity measurement is available, the peculiar velocity could also be estimated. This technique allows us to explore the natal kick velocity and the black hole formation mechanism \citep{miller14,russell15}. For a transient jet, high-cadence VLBI observations are needed to track the trajectories of the short-lived jet ejecta. These measurements potentially help to distinguish whether the radio flares associated with the state transition originate from the internal shock or the interplay between the jet and the surrounding interstellar medium \citep[e.g.][]{yang10,egron17}.
   
   AT2019wey (also known as ATLAS19bcxp, SRGA J043520.9+552226, SRGE J043523.3+552234) is an optical transient first discovered by the Asteroid Terrestrial-impact Last Alert System (ATLAS) survey \citep{tonry18} on 2019 December 7\footnote{\url{https://www.wis-tns.org//object/2019wey}}. On 2020 March 18, the X-ray telescopes ART-XC \citep{pavlinsky21} and eROSITA \citep{predehl21} on board the \textit{Spektrum-Roentgen-Gamma (SRG)} mission detected a previously uncatalogued X-ray source positionally coincident with the optical source \citep{mereminskiy20}. At first, based on the optical and X-ray properties, this source was tentatively classified as a BL Lac object (i.e. an active galactic nucleus at cosmological distance) in an active state \citep{lyapin20}. The lack of radio counterparts in the NRAO VLA Sky Survey \citep[NVSS,][]{condon98} and the Karl G. Jansky Very Large Array (VLA) Sky Survey \citep[VLASS,][]{lacy20} indicated its radio-weak nature. The 5-$\sigma$ flux density upper limits at 1.4 and 3~GHz are 2.4 and 0.7~mJy, respectively. We therefore conducted a one-hour VLA observation on 2020 May 27, and successfully detected a flat-spectrum radio source at the optical position, with a flux density of $\sim 200$~$\mu$Jy at C band \citep[$4-8$ GHz;][]{cao20}. In order to verify the existence of a compact radio core, we requested a follow-up European VLBI Network (EVN) observation. The EVN observation, scheduled at 6.7~GHz on 2020 October 17, detected a mJy-level radio source that was found to be much brighter than the source seen by the VLA \citep{giroletti20}. This indicated significant brightening. 
   
   The source was proposed as an accreting binary based on the detection of hydrogen lines at redshift $z = 0$ \citep{yao20a} and the detection in the radio bands \citep{yao20b}. Considering this, the radio brightening revealed by the EVN, and the radio spectral index change observed on 2020 August 2 with the VLA \citep{yao20b}, we applied multi-frequency Very Long Baseline Array (VLBA) observations to further unveil the nature of the EVN-detected component. Among the 68 stellar-mass black hole candidates currently known\footnote{\url{https://www.astro.puc.cl/BlackCAT/transients.php}}, 61 (including AT2019wey) are located at Galactic latitudes $|b|<10\degr$. Multi-frequency imaging observations at milliarcsecond (mas) scales allow us to probe the scatter-broadening effect caused by the intervening interstellar medium for these low-latitude sources. 
   We note that, in parallel with our efforts, \citet{yadlapalli21} conducted a VLBA experiment at 4.8~GHz, and detected (on 2020 September 9 and September 12) a resolved mJy-level radio source that they claimed is a compact jet. If the `radio core' nature is confirmed, it will extend the time baseline for probing the proper motion of the source.
   From the X-ray spectral and timing properties, \citet{yao20c} classified the source as a Galactic low-mass X-ray binary (LMXB), with the central accreting object being a black hole or neutron star. The authors of the detailed multi-wavelength study favoured the black-hole scenario and indicated that the companion is a short-period (P $\lesssim$ 16 h) low-mass ($<$1 M$_\sun$) star \citep{yao20d}. AT2019wey went through a large-amplitude variability in radio and X-rays from 2020 June to August, during which (i.e. on 2020 August 21) the source transitioned to the hard intermediate state (HIMS) from the LHS \citep{yao20c,yao20d}. The Galactic origin of the source was independently confirmed by \citet{mereminskiy21}.
   
   In this work, we report on the results of our three epochs of VLBI observations. Our observations and data reduction are described in Sect.~\ref{sec-2}. Our results are presented in Sect.~\ref{sec-3} and discussed in Sect.~\ref{sec-4}. Our conclusions are summarised in Sect.~\ref{sec-5}.

\section{Observations and data reduction}

\label{sec-2}

\subsection{Observations}

   \begin{table*}
      \caption[]{Observation information of the three VLBI experiments}
         \label{tab-1}
         \centering
         \begin{tabular}{c c c c c c c}
           \hline
           \noalign{\smallskip}
           Epoch & Obs. date & Array & Frequency & Bandwidth & Obs. time  & Antennas  \\
           &&&  {$[\mathrm{GHz}]$} & {$[\mathrm{MHz}]$} & {$[\mathrm{h}]$} & \\
           \noalign{\smallskip}
           \hline
           \noalign{\smallskip}
           1 & 2020-10-17 & EVN  &     6.7     &   256  & 4 & Jb Wb Ef Mc O8 T6 Tr Ys Ir Sr Km \\
           \noalign{\smallskip}
           2 & 2020-11-24 & VLBA & 1.6/4.5/6.7 & 384/256/256 & 4.5/1.5/1.5 & Br Fd Hn Kp La Mk Nl Ov Pt Sc   \\
           \noalign{\smallskip}
           3 & 2020-12-09 & VLBA & 1.6/4.5/6.7 & 384/256/256 & 4.5/1.5/1.5 & Br Fd Hn Kp La Mk Nl Ov Pt      \\
           \noalign{\smallskip}
           \hline
         \end{tabular}
         \begin{tablenotes}
         \item \emph{Notes}. Col. 4 -- observing frequency; Col. 5 -- total bandwidth per polarisation; Col. 6 -- total observing time. EVN telescope codes: Jb -- Jodrell Bank Mk2 (United Kingdom), Wb -- Westerbork (the Netherlands), Ef -- Effelsberg (Germany), Mc -- Medicina (Italy), O8 -- Onsala (Sweden), T6 -- Tianma (China), Tr -- Torun (Poland), Ys -- Yebes (Spain), Ir -- Irbene (Latvia), Sr -- Sardinia (Italy), Km -- Kunming (China); VLBA telescope codes: Br -- Brewster, Fd -- Fort Davis, Hn -- Hancock, Kp -- Kitt Peak, La -- Los Alamos, Mk -- Mauna Kea, Nl -- North Liberty, Ov -- Owens Valley, Pt -- Pie Town, Sc -- St. Croix.
         \end{tablenotes}
   \end{table*}

   The 6.7-GHz EVN experiment (project code: RSC07) was scheduled in the third disk-recording session of 2020, and lasted for 4~h. For most of the antennas, the data were recorded at a rate of 2~gigabits per second (Gbps), with two (left and right circular) polarisations, eight intermediate-frequency (IF) channels per polarisation, and 32~MHz bandwidth per IF. Westerbork station was limited to a recording rate of 1~Gbps, with four IFs per polarisation. The recorded data from each station were transferred to the Joint Institute for VLBI ERIC (JIVE) in Dwingeloo, the Netherlands, via the internet, and correlated with the EVN software correlator \citep[SFXC,][]{keimpema15}, with an integration time of 2s and 64 spectral channels per IF. 
   
   We obtained two epochs of multi-frequency (1.6, 4.5, and 6.7~GHz) VLBA observations (project code: BC270A/B). These can help to determine the spectral index and the possible time-evolving source position. At L band (1.6~GHz), the data were recorded at  a rate of 4~Gbps, with two polarisations, four IFs per polarisation, and 128~MHz bandwidth per IF. There are no useful data in the fourth IF due to the bandwidth limit of the 20cm receiver. This IF was therefore removed during the following data reduction, which results in a real data rate of 3~Gbps. Taking advantage of the broad 4 GHz bandwidth of the VLBA C-band receiver, the 4.5 GHz and 6.7 GHz observations were conducted simultaneously, each with 2~Gbps data rate and two IFs per polarisation. The data were correlated with the VLBA software correlator \citep[DiFX,][]{deller11} in Socorro, New Mexico, United States, with 2s integration time and 256 spectral channels per IF. The total observing time was 6~h per epoch, with 4.5~h used at 1.6~GHz. The aim was to detect any extended steep-spectrum radio emission if it exists at L band. The 1.5-h C-band observations were divided into three 30 min segments separated with a gap of 1 h to optimise the $(u,v)$ coverage. 
   
   The three VLBI experiments were carried out in phase-referencing mode \citep{beasley95}. The observation time was mostly spent on the target source (AT2019wey) and the nearby phase calibrator (J0442+5436, $1\fdg25$ away from the target). Most of the cycle times were 4~min, with 3~min spent on the target in each cycle. The bright and compact source DA193 was chosen as a fringe-finder. The target coordinates reported by the \textit{Gaia} Alerts\footnote{\url{http://gsaweb.ast.cam.ac.uk/alerts/home}} were taken as the initial position for VLBI pointing in the EVN experiment, and the accurate coordinates measured with the EVN were further used for the subsequent VLBA experiments. We also included J0418+5457 in the VLBA C-band observations, a radio quasar in the third realisation of the International Celestial Reference Frame \citep[ICRF3;][]{charlot20}, for position-checking purposes. The coordinates of the three selected calibrators were taken from the radio fundamental catalogue (rfc\_2020c) of the Astrogeo database\footnote{\url{http://astrogeo.org/calib/search.html}}. The observation information is summarised in Table~\ref{tab-1}.
   
   \subsection{Data reduction}
   
   \begin{figure}[!htb] 
   \centering
   \includegraphics[width=0.45\textwidth]{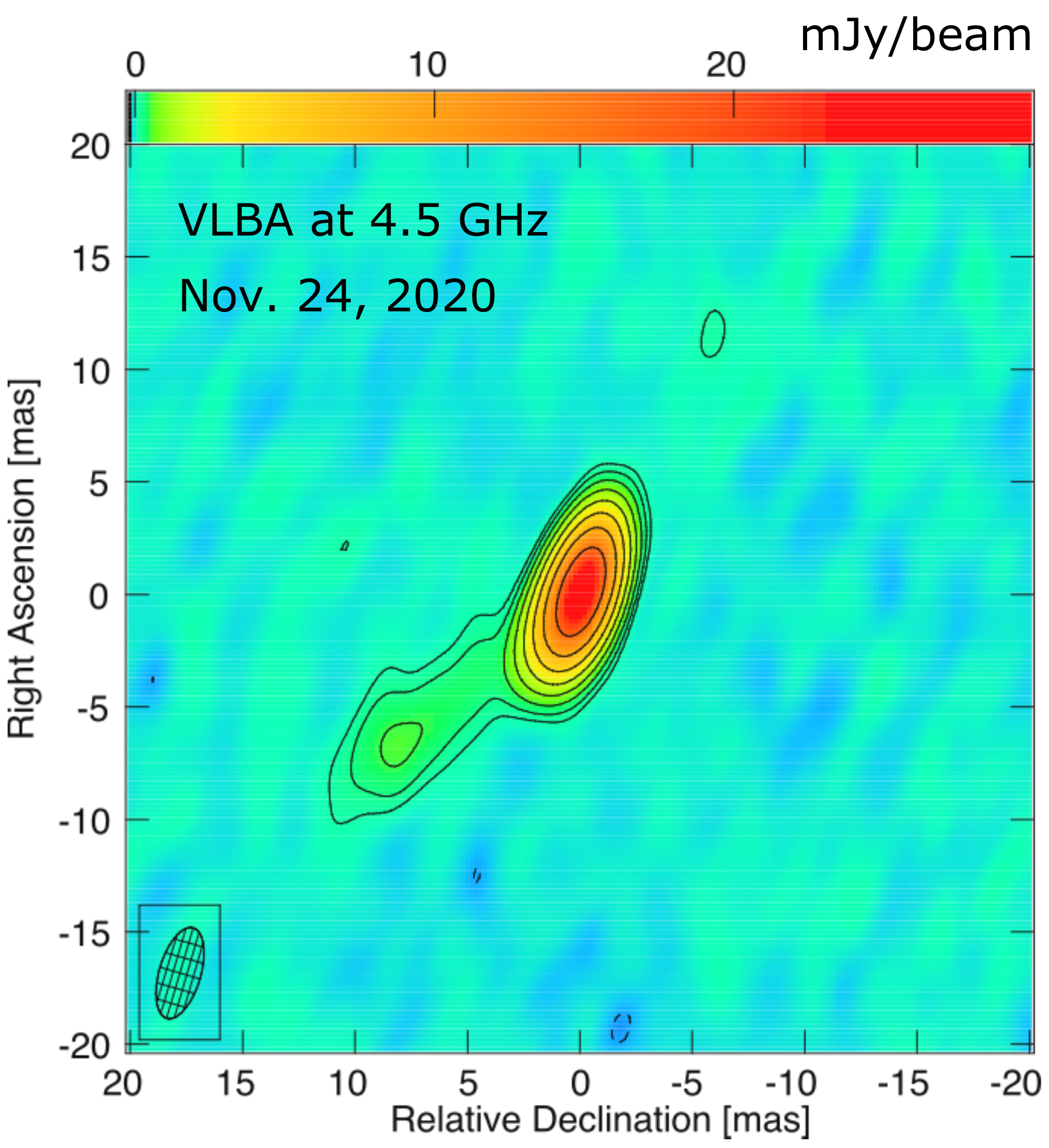}
   \caption{Naturally weighted \textsc{clean} image of the phase calibrator J0442+5436 at 4.5~GHz. The peak brightness is 29.8~mJy\,beam$^{-1}$. The 1-$\sigma$ image noise level is 41.4~$\mu$Jy\,beam$^{-1}$. The first contours are at $\pm 3 \sigma$, and the positive contours increase by a factor of two. The synthesised beam shown in the lower left corner is 1.8~mas $\times$ 4.2~mas with the position angle of the major axis at $-16\fdg3$.}
              \label{fig-1}%
    \end{figure}

   The correlated EVN data were calibrated in the NRAO Astronomical Processing System \citep[AIPS;][]{greisen03}, generally following the EVN data reduction Guide\footnote{\url{https://www.evlbi.org/evn-data-reduction-guide}}.
   A priori amplitude calibration was done using the system temperatures recorded during the observations at the VLBI stations and the antenna gain curves. As the system temperatures were not available for the Kunming telescope, the nominal system equivalent flux density (SEFD) was used instead. Parallactic angle and ionospheric corrections were then performed. After manual phase calibration carried out on the fringe-finder source DA193 to remove the instrumental delays, global fringe-fitting was performed on DA193 and the phase calibrator J0442+5436. Bandpass corrections were determined using DA193. 
   
   The VLBA data calibration followed the AIPS Cookbook\footnote{\url{http://www.aips.nrao.edu/cook.html}}. Compared with the EVN case, some additional steps were added: the three-frequency data were first separated and fixed; the (likely) inaccurate Earth orientation parameters and the sampler biases were also corrected. After manual phase calibration, the 4.5 GHz and 6.7 GHz data were isolated using the task \textsc{uvcop}. Moreover, the a priori amplitude calibration was postponed until the bandpass solutions were available.  
   
   The calibrators were imaged in \textsc{Difmap} \citep{shepherd97}. In order to mitigate the residual phase errors caused by the source structure, global fringe-fitting was rerun taking into account the brightness distribution model of the phase calibrator, which does show an extended structure (Fig.~\ref{fig-1}). Gain scale factors deviating by more than $\pm 5\%$ from unity, which is typical for the EVN antennas, were applied to the visibilities in AIPS using \textsc{clcor}. Finally, the calibrated visibility data for the sources were exported to and analyzed in \textsc{Difmap}.
    
\section{Results} 

\label{sec-3}
   
   \begin{figure*}[!htb] 
   \centering
   \includegraphics[width=0.99\textwidth]{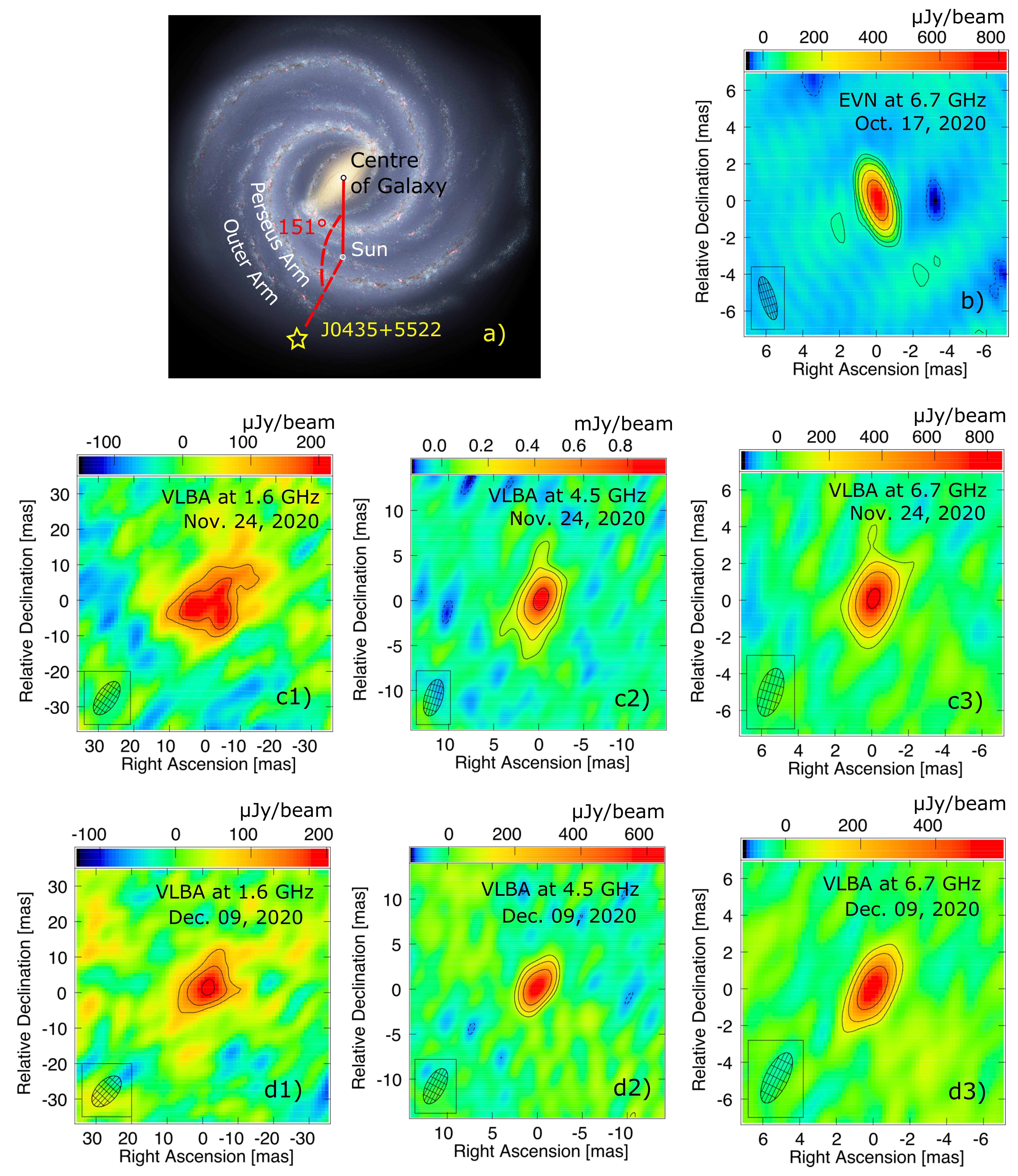}
   \caption{Position in the Galaxy and multi-frequency, multi-epoch VLBI images of the target source AT2019wey. The rough position of AT2019wey (J0435+5522) in the Galaxy is shown in panel (a) (background image credit: NASA / JPL-Caltech / R. Hurt [SSC/Caltech]).
   From panel (b) to (d3), there are the naturally weighted VLBI images of AT2019wey made with the \textsc{modelfit} program in \textsc{Difmap}. The synthesised beams (shown in the lower left corners), peak brightnesses, and 1-$\sigma$ image noise levels are listed in Table~\ref{tab-2}. The first contours are at $\pm 3 \sigma$, and the positive contours increase by a factor of $\sqrt{2}$ for 1.6 GHz, and by a factor of 2 for 4.5 GHz and 6.7 GHz frequencies.}
              \label{fig-2}%
    \end{figure*}

   To maximise the sensitivity, natural weighting was used for imaging the target AT2019wey. The source was clearly detected at all epochs and frequencies. We used a single circular Gaussian component fitted to the calibrated visibility data to model the source brightness distribution. The phase-referenced images produced in \textsc{Difmap} are shown in Fig.~\ref{fig-2}. The image and source parameters are listed in Tables~\ref{tab-2} and \ref{tab-3}, respectively. We used the method provided by \citet{lee08} to estimate the errors of the flux densities and source sizes, which were then propagated to the brightness temperatures. For the flux densities, systematic errors (we conservatively took 10\% of the measured values) were also included to account for the imperfect amplitude calibration.
   
   The minimum resolvable size $\theta_{\rm min}$ of an interferometer, which can be smaller than the beam size (i.e. the diffraction limit) measured at the full width at half maximum (FWHM) of the synthesised beam, is calculated as
   \begin{equation}
       \theta_{\rm min} = b_{\psi}\sqrt{\frac{4\,{\rm ln\,2}}{\pi}{\rm ln}\left(\frac{S/N}{S/N-1}\right)}
       \label{eq-1}
   ,\end{equation}
   \citep{kovalev05}, where $b_{\psi}$ is the beam size along the direction at a position angle $\psi$ for which the resolution limit is determined. $b_{\psi}=\sqrt{b_{\rm max} \times b_{\rm min}}$ was used for our estimation ($b_{\rm max}$ and $b_{\rm min}$ are the beam major and minor axis FWHM, respectively). The signal-to-noise ratio (S/N) is calculated as the peak brightness divided by the r.m.s. (see Table~\ref{tab-2}). As indicated in Table~\ref{tab-3}, the source sizes are all larger than the corresponding minimum resolvable sizes, suggesting the source is resolved by the VLBI observations.
   
   For a Galactic radio source with a circular Gaussian brightness distribution, the brightness temperature $T_{\rm b}$ is derived as
   \begin{equation}
       T_{\rm b} = 1.22 \times 10^{12} \frac{S_{\nu}}{\theta^2\nu^2} ~ [{\rm K}]
       \label{eq-2}
   \end{equation}
   \citep{condon82,lee08}, where $S_{\nu}$ is the flux density in Jy, $\theta$ is the source size in mas, and $\nu$ is the observing frequency in GHz. The inferred brightness temperatures are larger than $10^6$~K (Table~\ref{tab-3}), which is consistent with the synchrotron emission origin of the source radio emission.
   
   Taking 1~kpc as the lower limit of the source distance ($d$), which is estimated based on the Galactic extinction \citep{yao20d}, the lower limit of the monochromatic luminosity $L_{\nu} = 4\pi\,d^2\,S_{\nu}$ can then be constrained (Table~\ref{tab-3}). The value of $\sim 10^{11}$~W\,Hz$^{-1}$ is at the lower end of the luminosity range of the previously known BHXBs \citep{gallo06,yao20d}.
    
   \begin{table*}
      \caption[]{Image parameters of AT2019wey}
         \label{tab-2}
         \centering
         \begin{tabular}{c c c c c c c}
           \hline
           \noalign{\smallskip}
           Epoch & $\nu$ [GHz]  & Beam [mas $\times$ mas] & P.A. $[\degr]$ & Peak br. [mJy\,beam$^{-1}$] & r.m.s. [$\mu$Jy\,beam$^{-1}$] &  Fig. \\
           \noalign{\smallskip}
           \hline
           \noalign{\smallskip}
           1 & 6.7 & 2.4 $\times$ 0.8  & 16.7    & 0.82  & 9.6  & \ref{fig-2}b  \\
           \noalign{\smallskip}
           2 & 1.6 & 10.7 $\times$ 5.5 & $-34.2$ & 0.22  & 38.2 & \ref{fig-2}c1 \\
           \noalign{\smallskip}
             & 4.5 & 4.2 $\times$ 2.0  & $-16.7$ & 0.96  & 32.4 & \ref{fig-2}c2 \\
           \noalign{\smallskip}
             & 6.7 & 2.7 $\times$ 1.3  & $-16.6$ & 0.85  & 31.9 & \ref{fig-2}c3 \\ 
           \noalign{\smallskip}
           3 & 1.6 & 10.5 $\times$ 5.9 & $-42.2$ & 0.20  & 31.5 & \ref{fig-2}d1 \\
           \noalign{\smallskip}
             & 4.5 & 4.3 $\times$ 2.1  & $-27.5$ & 0.65  & 27.7 & \ref{fig-2}d2 \\
           \noalign{\smallskip}
             & 6.7 & 3.0 $\times$ 1.2  & $-26.4$ & 0.58  & 29.5 &  \ref{fig-2}d3 \\
           \hline
         \end{tabular}
         \begin{tablenotes}
         \item \emph{Notes}. The source images are shown in Fig.~\ref{fig-2}. Col. 2 -- observing frequency; Col. 3 -- restoring beam size, i.e., the FWHM of the synthesised beam, represented by the beam major axis ($b_{\rm max}$) $\times$ minor axis ($b_{\rm min}$); Col. 4 -- position angle of the beam major axis, measured from north through east; Col. 5 -- peak brightness; Col. 6 -- 1-$\sigma$ image noise level; Col. 7 -- figure number. 
         \end{tablenotes}
   \end{table*} 
    %
   
   \begin{table*}
      \caption[]{Source parameters of AT2019wey}
         \label{tab-3}
         \centering
         \begin{tabular}{c c c c c c c c}
           \hline
           \noalign{\smallskip}
           Epoch & $\nu$ [GHz]  & $S_{\nu}$ [mJy] & $\theta$ [mas] & $\theta_{\rm min}$ [mas] & $T_{\rm b}$ [$10^7$K] & $L_{\nu}$ [$10^{11}$ W\,Hz$^{-1}$] & Fig. \\
           \noalign{\smallskip}
           \hline
           \noalign{\smallskip}
           1 & 6.7 & 1.2 $\pm$ 0.2  & 0.7 $\pm$ 0.1 & 0.14  & 6.4 $\pm$ 1.8 & > 1.4 & \ref{fig-2}b  \\
           \noalign{\smallskip}
           2 & 1.6 & 1.2 $\pm$ 0.6  & 16.6 $\pm$ 7.6      & 3.19  & 0.21$^{+0.84}_{-0.16}$ & > 1.5 & \ref{fig-2}c1 \\
           \noalign{\smallskip}
             & 4.5 & 1.5 $\pm$ 0.4  & 2.0 $\pm$ 0.4   & 0.50  & 2.2 $\pm$ 1.0 & > 1.8 & \ref{fig-2}c2 \\
           \noalign{\smallskip}
             & 6.7 & 1.2 $\pm$ 0.3  & 1.0 $\pm$ 0.2   & 0.34  & 2.9 $\pm$ 1.4 & > 1.4 & \ref{fig-2}c3 \\ 
           \noalign{\smallskip}
           3 & 1.6 & 0.8 $\pm$ 0.3  & 13.8 $\pm$ 5.9     & 3.07  & 0.18$^{+0.62}_{-0.13}$ & > 0.9 & \ref{fig-2}d1 \\
           \noalign{\smallskip}
             & 4.5 & 0.8 $\pm$ 0.2  & 1.4 $\pm$ 0.3   & 0.59  & 2.5 $\pm$ 1.3 & > 1.0 &\ref{fig-2}d2 \\
           \noalign{\smallskip}
             & 6.7 & 0.8 $\pm$ 0.2  & 0.9 $\pm$ 0.2   & 0.41  & 2.4 $\pm$ 1.3 & > 0.9 & \ref{fig-2}d3 \\
           \hline
         \end{tabular}
         \begin{tablenotes}
         \item \emph{Notes}. The model parameters corresponding to the source images in Fig.~\ref{fig-2}. Col. 2 -- observing frequency; Col. 3 -- flux density; Col. 4 -- angular size, i.e. the FWHM of the circular Gaussian component; Col. 5 -- minimum resolvable angular size of the VLBI array; Col. 6 -- brightness temperature;  Col. 7 -- lower limit of the monochromatic luminosity, calculated by assuming that the source is located 1~kpc away from the Earth; Col. 8 -- figure number. The errors of brightness temperatures at 1.6~GHz were likely overestimated and thus determined with the lower and upper bounds of the flux densities and source sizes.
         \end{tablenotes}
   \end{table*}  
    
   The target appears more compact at C band, with the source sizes being comparable to the beam minor axes (Tables~\ref{tab-2} and \ref{tab-3}). The phase-referenced coordinates were then measured at this band with the task \textsc{tvmaxfit} in AIPS (Table~\ref{tab-4}). The target source positions relative to the average position (RA = $ 04^{\rm h} 35^{\rm m} 23\fs2733830$, Dec = $55\degr 22\arcmin 34\farcs289098$) are shown in Fig.~\ref{fig-3}. From the right panel of Fig.~\ref{fig-3}, a random scatter of our five measurements can be seen.
   Therefore, we use the standard deviations of the measured coordinates ($\sigma_{\mathrm{RA}} = 0.28$~mas, $\sigma_{\mathrm{Dec}} = 0.07$~mas; Fig.~\ref{fig-3}) to represent the target position uncertainty. This is consistent with the uncertainties expected from the conventional phase-referencing astrometry with $1\fdg25$ target--calibrator separation \citep{rioja21}. The check source (J0418+5457) has an obvious position offset of $2-3$~mas from its phase centre. However, the angular separation of the check source from the phase-reference calibrator (J0442+5436) is larger: $3\fdg48$. In relative astrometry, this offset is sufficient ($<10$~mas) to avoid introducing an extra inconstant position error term resulting from the phase calibrator \citep{reid14}. 
   
   
   \begin{table}
      \caption[]{Phase-referenced coordinates of AT2019wey}
         \label{tab-4}
         \centering
         \begin{tabular}{c c c c}
         \hline
           \noalign{\smallskip}
           Epoch & $\nu$ [GHz]  & RA [h m s] & Dec [\degr\,\arcmin\,\arcsec] \\
           \noalign{\smallskip}
           \hline
           \noalign{\smallskip}
           1 & 6.7 & 04 35 23.2733809 & 55 22 34.289112 \\
           \noalign{\smallskip}
           2 & 4.5 & 04 35 23.2733556 & 55 22 34.289194 \\
           \noalign{\smallskip}
             & 6.7 & 04 35 23.2733787 & 55 22 34.289075 \\
           \noalign{\smallskip}
           3 & 4.5 & 04 35 23.2734046 & 55 22 34.289099 \\
           \noalign{\smallskip}
             & 6.7 & 04 35 23.2733950 & 55 22 34.289009 \\
           \noalign{\smallskip}
           \hline
         \end{tabular}
         \begin{tablenotes}
         \item \emph{Notes}. Col. 2 -- observing frequency; Col. 3 -- right ascension (J2000); Col. 4 -- declination (J2000).
         \end{tablenotes}
   \end{table}  
   
\section{Discussion}

\label{sec-4}
   
   \subsection[Radio properties of AT2019wey]{Radio properties of \object{AT2019wey}}
   
   \begin{figure*}[!htb] 
   \centering
   \includegraphics[width=0.9\textwidth]{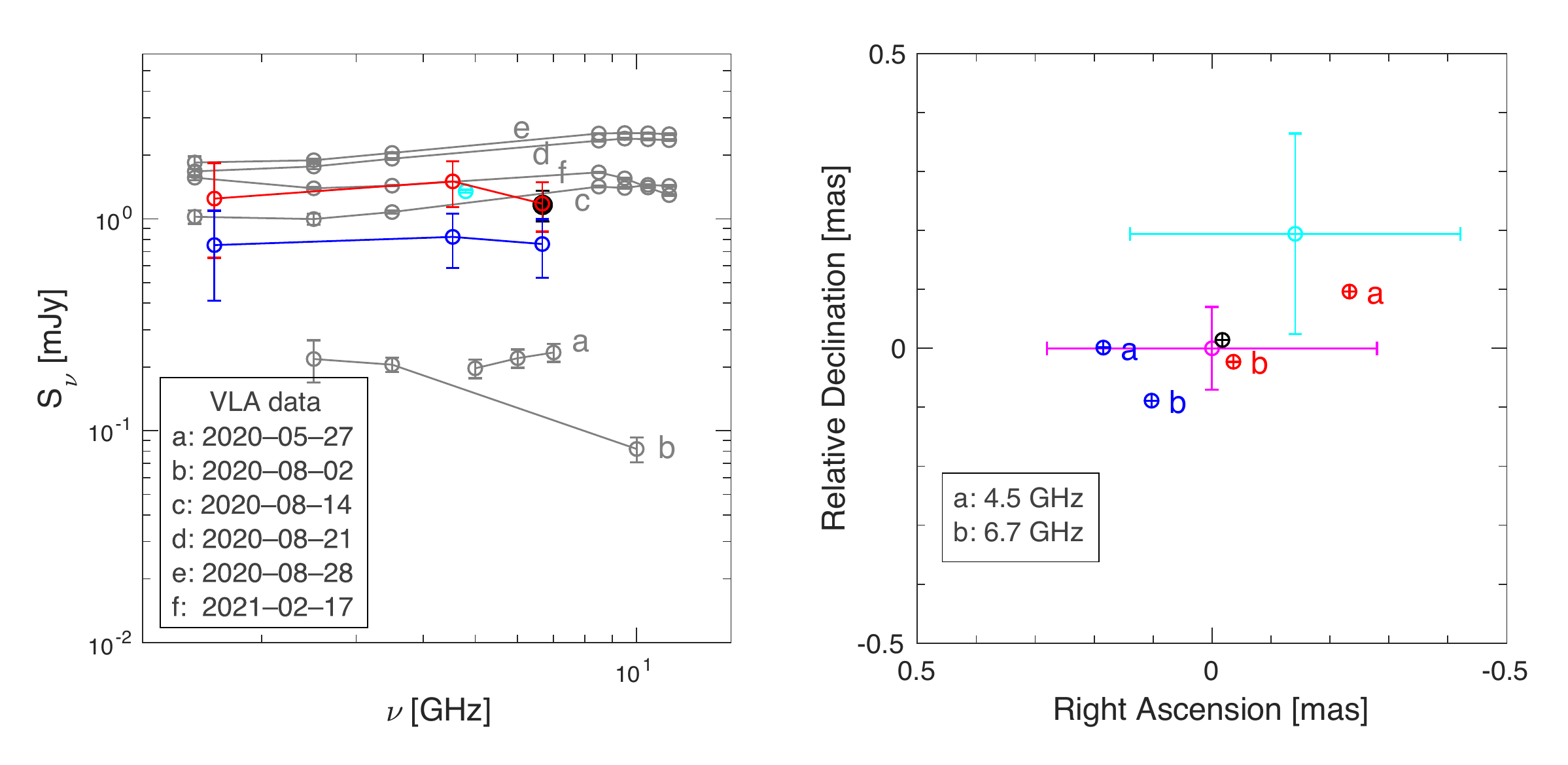}
   \caption{Radio spectrum characteristics and phase-referenced positions of AT2019wey. Black: EVN epoch, red: VLBA epoch 1, blue: VLBA epoch 2, cyan: measurements by \citet{yadlapalli21} with the VLBA at 4.8~GHz. Left panel: Flux density vs. frequency. The lines connecting the data points are not fitted curves but are simply to guide the eye. The flux densities measured by the VLA (grey; \citealt{yao20d}) are also shown for comparison, and the observing dates are denoted by lowercase letters a $-$ f accordingly. Right panel: Relative astrometric positions with the average position (magenta; obtained by our five measurements listed in Table \ref{tab-4}) as the reference point. The lowercase letters a and b correspond to 4.5 and 6.7~GHz, respectively.}
              \label{fig-3}%
    \end{figure*}
   
   According to our three-epoch VLBI observations, only one component was significantly detected in AT2019wey. The flux densities measured at the three different frequencies in the two VLBA experiments are consistent with each other within their uncertainties, indicating a flat radio spectrum (spectral index $\alpha \approx 0$; $S_{\nu} \propto \nu^\alpha$; see Table~\ref{tab-3} and Fig.~\ref{fig-3}). The random position distribution of the target (Fig.~\ref{fig-3}) strongly disfavours the detection of the source motion on the sky. This, together with the flat radio spectrum, supports the hypothesis that the single VLBI component we detected is a compact jet rather than a moving knot ejected in relation to
   the radio brightening in the summer of 2020. After correcting the position offset shown by the check source (the phase calibrator used by \citealt{yadlapalli21}) with respect to its phase centre, the target position given in \citet{yadlapalli21} is consistent with that measured by us (Fig.~\ref{fig-3}). This strengthens the non-detection of the source motion, even over a three-month time baseline. However, we note that the non-point source characteristic of the phase calibrator (Fig.~\ref{fig-1}) and once again the large target--calibrator angular separation ($3\fdg48$) could influence the accurate link of the two reference frames. 
   
   A flat-spectrum mJy-level radio source was detected by the VLA on 2021 February 17 (Fig. \ref{fig-3}; \citealt{yao20d}). Therefore, the current available radio data (Fig. \ref{fig-3}) suggest that the source remained bright at mJy level from mid-August of 2020 to mid-February of 2021, which is consistent with the `hard-only' scenario proposed by \citet{yao20c,yao20d}. The flux densities and the source sizes at the second VLBA epoch appear systemically lower than that at the first epoch (see Fig.~\ref{fig-3} and \ref{fig-4}), which implies ongoing radio variability. From the monitoring conducted with the Neutron star Interior Composition Explorer (NICER) experiment \citep{gendreau12}, the source also showed high and variable X-ray flux until April 2021, and did not show any signature of falling back from the HIMS to LHS \citep{mereminskiy21}. Such a long outburst may facilitate a quasi-simultaneous monitoring to explore the correlation between the radio and X-ray flux density variations.
   
   \subsection{Source distance and peculiar velocity}
   
\begin{figure*}[!htb] 
   \centering
   \includegraphics[width=0.9\textwidth]{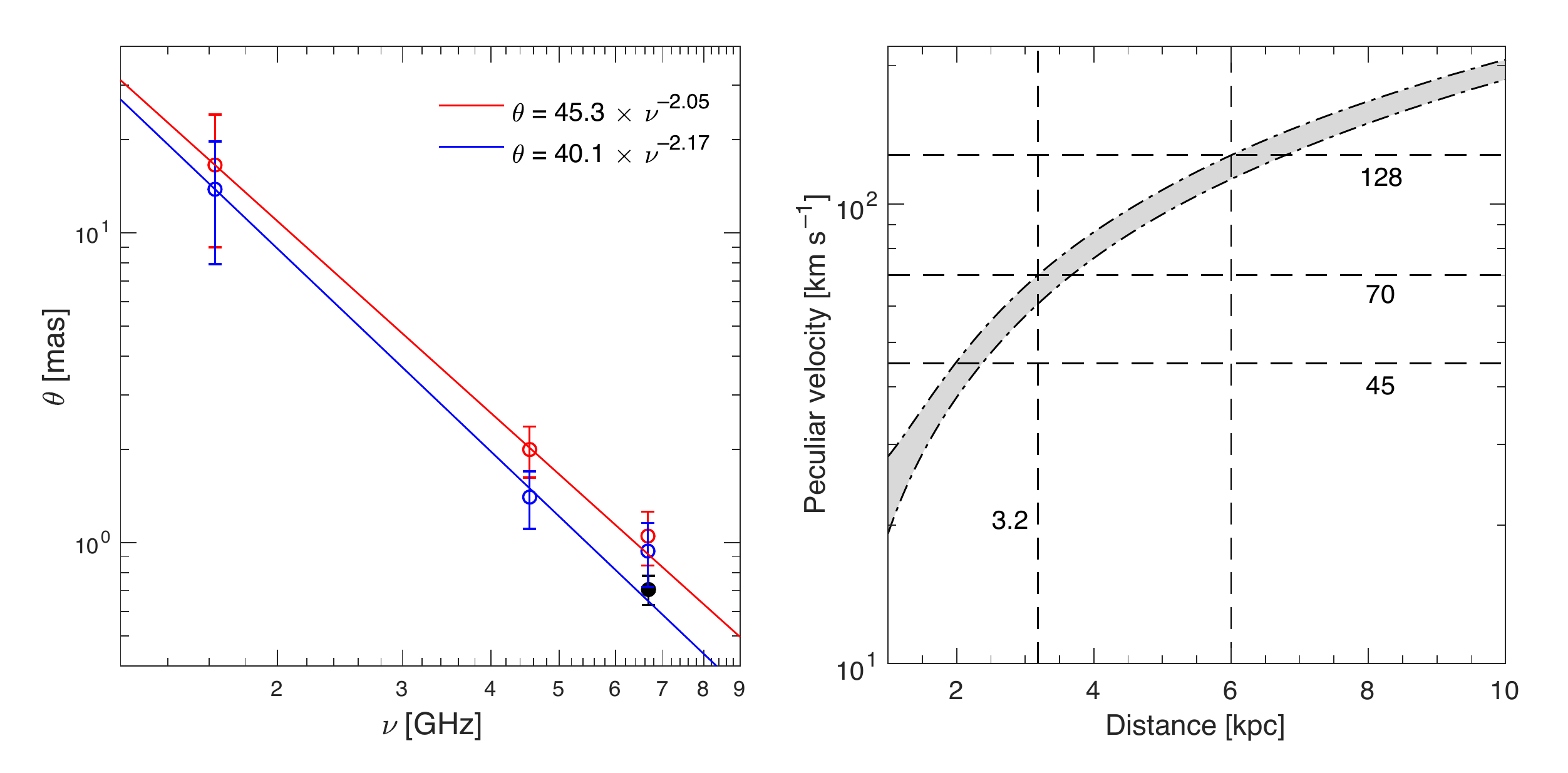}
   \caption{Angular broadening and peculiar velocity of AT2019wey. Left panel: Source size vs. frequency ($\theta = A \times \nu^\beta$). The 68\% confidence intervals of the coefficient $A$ and exponent $\beta$ are: [$43.9$, $46.7$] and [$-2.12$, $-1.99$] for VLBA epoch 1; and [$35.1$, $45.1$] and [$-2.43$, $-1.92$] for VLBA epoch 2, respectively. Black: EVN epoch, red: VLBA epoch 1, blue: VLBA epoch 2.  Right panel: Upper limit on the peculiar velocity of the source set by our proper motion
measurement shown by the shaded area. We assumed zero radial velocity, and took into account the errors of the Galactic parameters. The velocities of the stars in the Galactic disc are typically lower than 45~km\,s$^{-1}$ \citep{mignard00}, and 70~km\,s$^{-1}$ is the maximum recoil velocity resulting from typical Blaauw kicks in LMXBs \citep{nelemans99}.}
              \label{fig-4}%
    \end{figure*}

   The source is at approximately the Galactic anticentre direction (its Galactic longitude and latitude are $l = 151\fdg16116$ and $b = 5\fdg29985$, respectively)\footnote{\url{https://ned.ipac.caltech.edu/coordinate_calculator}}, which sets an upper limit on the source distance of $d < 10$~kpc \citep{yao20d}. The low Galactic latitude motivates us to explore the possibility of angular broadening caused by the ionised interstellar medium in the Galaxy. As shown in Fig.~\ref{fig-4}, the source sizes and frequencies closely follow the relation $\theta \propto \nu^{-2}$, which suggests that angular broadening indeed affects the measured sizes of the intrinsically compact radio source.
   The phase calibrator does not show clear signatures of scattering, implying the angular broadening is most likely caused by scattering medium very near the target source. This was also proposed to explain the `fuzzy' radio core seen in MAXI J1659--152 by \citet{paragi13}, where the scatter broadening is not well established due to the lack of multi-frequency data. If the scattering medium is located in the Perseus arm or in the more distant Outer arm, then the source distance would be $\sim 6-10$~kpc, inferred from the spiral-arm model given in \citet{bobylev14}. In this case, the 5 GHz radio luminosity of the source would be larger than 10$^{22}$ W, which is close to the typical values of the known BHXBs in hard state \citep{yao20d}.
   
   The position uncertainty of the source can set an upper limit on the proper motion: $\mu_\alpha= 3.8$ mas yr$^{-1}$ and $\mu_\delta=0.9$ mas yr$^{-1}$, which correspond to a space velocity of $< 180$ km s$^{-1}$ in right ascension and $< 40$ km s$^{-1}$ in declination, respectively, for a source distance of less than 10 kpc. Even for the most distant case (i.e. 10 kpc), and a slow motion with a velocity of $0.1 c$, the inferred proper motion of $\sim 1.7$ mas day$^{-1}$ for an ejected knot is still much larger than the estimated proper motion upper limit. Therefore, the VLBI-detected component is unlikely a jet knot. Assuming that the radial velocity ---which has not yet been measured--- is equal to zero, we calculated the upper limit of the source peculiar velocity in the local standard of rest (LSR) frame as shown in Fig.~\ref{fig-4}\footnote{\url{https://idlastro.gsfc.nasa.gov/ftp/pro/astro/gal_uvw.pro}} (\citealt{johnson87,russell15}). The Sun's distance from the Galactic centre of $8.2 \pm 0.1$ kpc, the Sun's velocity relative to LSR of ($10 \pm 1$, $11 \pm 2$, $7 \pm 0.5$) km s$^{-1}$, and a flat rotation curve with a circular velocity of 238 km s$^{-1}$ were adopted from \citet{bland-Hawthorn16}. If the source is located at a distance of $\gtrsim 3$ kpc, the upper limit on the peculiar velocity is then greater than 70 km s$^{-1}$. This is compatible with the existence of an asymmetric natal kick \citep{russell15}. Otherwise, a non-zero radial velocity is needed. Our proper motion measurement is based on a 53-day time baseline. For such a long-term radio outburst as seen from AT2019wey, more stringent constraint of the peculiar velocity can be anticipated from high-precision relative astrometry conducted with the VLBI monitoring for only one single outburst. Better constraints on the source distance and radial velocity will help in this respect.
   
    \subsection{Jet parameters}
   
   The brightness temperatures measured at different frequencies are expected to be close to each other according to the model for a compact jet, where the source angular size and frequency scale as $\theta \propto \nu^{-1}$ \citep{blandford79,pushkarev15}. However, the brightness temperatures measured at L band appear an order of magnitude lower than that measured at C band (Table \ref{tab-3}). This discrepancy can be understood in the context of interstellar scattering, which becomes stronger towards lower frequencies. 
   
   Based on the parameter-estimation method \citep{fender06,longair11}, angular broadening can result in overestimation of the jet minimum energy and mean power, and underestimation of the magnetic field strength. The single-peaked hydrogen line profile and the X-ray reflection spectrum imply that the source is viewed at low inclination (i.e. close to face on; \citealt{yao20d}). This is also supported by the observed radio variability and flat radio spectrum. Assuming the jet Doppler factor $\delta \gtrsim 1$, and the magnetic field and particle energy densities are in equipartition, then the upper limit on the intrinsic source size $\theta_{\rm in} \approx 20-100$ $\mu {\rm as}$ (here we take the equipartition brightness temperature $T_{\rm eq} = 5 \times 10^{10}$ K, and $\delta = T_{\rm b}/T_{\rm eq}$; \citealt{readhead94}), which is clearly lower than that measured by VLBI (Table \ref{tab-3}).
   
\section{Conclusions}

\label{sec-5}

   Our three-epoch VLBI observations of AT2019wey spanning about 2 months reveal a flat-spectrum ($\alpha \approx 0$) radio component and do not detect significant apparent proper motion. These, together with the radio variability revealed by the two VLBA epochs, provide evidence that the VLBI-detected component is a compact jet. The brightening in the radio and X-rays in the summer of 2020 suggests a transformation of the disk--corona structure into a more efficient state for producing a jet. The source stayed bright in the radio, implying that the conditions for the formation of a powerful jet remained unchanged. This is also supported by the X-ray monitoring by NICER and the most recent VLA observation on 2021 February 17. No ejecta were seen in our VLBI observations.
     
   The source is at a low Galactic latitude ($b \sim 5\degr$). The change of its angular size with observing frequency suggests angular broadening by the intervening ionised Galactic interstellar medium. If the scattering medium is from a Galactic spiral arm, the lower limit on the source distance would be 6~kpc. The upper limit on the source peculiar velocity is fully compatible with the scenario of an asymmetric natal kick, if the source distance is larger than 3.2~kpc.
     
   The brightness temperature of $T_{\rm b} \sim 10^{7}$~K indicates that the radio emission has a non-thermal synchrotron origin. For the weak radio sources like AT2019wey, lower frequency observations are necessary, where better sensitivity can be achieved. Multi-frequency VLBI observations are therefore needed to unveil the possible scatter-broadening effect for the sources at low Galactic latitudes, which may lead to overestimation of the source angular size and thus affect estimations of the jet parameters.

\begin{acknowledgements}
      The EVN is a joint facility of independent European, African, Asian and North American radio astronomy institutes. Scientific results from data presented in this publication are derived from the following EVN project code: RSC07. The Very Long Baseline Array is a facility of the National Science Foundation operated under cooperative agreement by Associated Universities, Inc. HMC acknowledges support by the National Natural Science Foundation of China (Grants No. U2031116 and U1731103). SF and K\'EG thank the Hungarian National Research, Development and Innovation Office (OTKA K134213 and 2018-2.1.14-T\'ET-CN-2018-00001) for support.
\end{acknowledgements}

%
%

\end{document}